\documentclass[preprint,preprintnumbers,amsmath,amssymb,aps]{revtex4}
\usepackage{amsfonts}
\usepackage{amssymb}
\usepackage{latexsym}
\usepackage{graphicx}
\usepackage{psfig}

\begin{document}


\title{Shannon entropies of atomic structure factors, off-diagonal order and electron correlation}
\author{Robin P. Sagar  and Nicolais L. Guevara \footnote{Present address: 
 Instituto de Ciencias Nucleares, Universidad Nacional Aut\'onoma de M\'exico, D.F., 04510 M\'exico. }
   } 
\affiliation{Departamento  de  Qu\'{\i}mica,  Universidad  Aut\'onoma
Metropolitana Apartado Postal 55-534, Iztapalapa, 09340 M\'exico D.F., M\'exico}
\date{\today}


\begin{abstract}
Shannon entropies of one- and two-electron atomic structure factors in the position and momentum representations are used to examine the behavior of the off-diagonal elements of density matrices with respect to the uncertainty principle and to analyze the effects of electron correlation on  off-diagonal order. We show that electron correlation induces off-diagonal order in position space which is characterized by larger entropic values. Electron correlation in momentum space is characterized by smaller entropic values as information is forced into regions closer to the diagonal.
Related off-diagonal correlation functions are also discussed.
\end{abstract}
\maketitle
\newpage




\section{Introduction }

A fundamental difference between classical and quantum mechanics is the appearance of the wave-particle duality in the latter. The usual language used to distinguish between the two is that of decoherence-coherence with the former implying a transition to the classical regime. In atomic structure studies, the terminology used is that of electronic localization-delocalization, normally employed in the description of electronic charge densities, in the position or momentum representations. It is thought that the extent of this localization-delocalization behavior is due in part to the effects of electron correlation.  These quantum effects manifest themselves in a variety of physical phenomena such as atomic shell structure, chemical binding and reactivity, and superconductivity in condensed matter systems.

The wave-particle duality in quantum mechanics finds it expression in the Heisenberg uncertainty principle, i.e. it is impossible to localize the position of an electron without delocalizing its momentum. There are various formulations of the uncertainty principle but the one which will interest us here is the information theoretic one at the one-electron level \cite{bbm,gadre,guevarajcp1},
\begin{equation}
 0 \leq 3(1+ \ln \pi) \leq S_{\rho}^{u}+ S_{\pi}^{u}=S_{t}^{u}.  
\label{guevara03}
\end{equation}
In the above, $S_{\rho}^{u}$ and $S_{\pi}^{u}$ are the unity-normalized one-electron Shannon entropies in position and momentum space respectively, defined as 
\begin{equation}
S_{\rho}^{u}= - \int \frac{\rho({\bf r})}{N} \ln {\Bigg [} \frac{\rho({\bf r})}{N}{\Bigg ]} d {\bf r}
\label{shanr}
\end{equation}
\begin{equation}
S_{\pi}^{u}= - \int \frac{\pi({\bf p})}{N} \ln {\Bigg [}\frac{\pi({\bf p})}{N}{\Bigg ]} d {\bf p}.
\label{shanp}
\end{equation}
$\rho(\bf r)$ and $\pi(\bf p)$ are the one-electron charge and momentum densities, and both 
$\pi(\bf p)$ and $\rho(\bf r)$ are normalized to N, the number of electrons in the system. 
These densities provide  the local information about the system at the one-electron level in each space. 

The spin-free one-electron reduced density matrix is defined in terms of the full N-order density matrix \cite{davidson} in position space as
\begin{equation}
\gamma ({\bf r_1};{\bf r_1}^{\prime}) = N \int
\Psi ({\bf x}_1, \cdots,{\bf x}_N)
\Psi^* ({\bf x}_1^{\prime}, \cdots,{\bf x}_N)
\delta(\sigma _1 - \sigma^{\prime} _1)d{\sigma}_1d{\sigma^{\prime}}_1 
d{\bf x}_2 \cdots d{\bf x}_N
\label{ODM1}
\end{equation}
and in momentum space as
\begin{equation}
\hat{\gamma}({\bf p_1};{\bf p_1}^{\prime}) = N \int
\Phi ({\bf y}_1, \cdots,{\bf y}_N)
\Phi^* ({\bf y}_1^{\prime}, \cdots,{\bf y}_N)
\delta(\sigma _1 - \sigma^{\prime} _1)d{\sigma}_1d{\sigma^{\prime}}_1 
d{\bf y}_2 \cdots d{\bf y}_N,
\label{ODM2}
\end{equation}
where $\Psi$ and $\Phi$ are the N-electron wavefunctions in position and momentum space. Combined space-spin co-ordinates, ${\bf x}_i=({\bf r}_i,\sigma _i)$, and momentum-spin co-ordinates, ${\bf y}_i=({\bf p}_i,\sigma _i)$, are used.

The two representations are connected by a Dirac-Fourier transformation
\begin{equation}
\hat{\gamma} ({\bf p};{\bf p}^{\prime}) = (2\pi)^{-3} \int \gamma ({\bf r};{\bf r}^{\prime})
\exp[-i({\bf p} \cdot {\bf r} - {\bf p}^{\prime} \cdot {\bf r}^{\prime})] d{\bf r} d{\bf r}^{\prime}.
\label{ODMFT}
\end{equation}
%
%
The one-electron densities in each space are the diagonal elements of the one-electron reduced density matrices  (${\bf r_1} = {\bf r_1}^{\prime}$ and $ {\bf p_1} = {\bf p_1}^{\prime}$) , i.e. $\rho({\bf r})=\gamma ({\bf r_1};{\bf r_1})$ and $\pi({\bf p})=\hat{\gamma} ({\bf p_1};{\bf p_1})$. The off-diagonal elements, $ {\bf r_1} \neq {\bf r_1}^{\prime}$ and $ {\bf p_1} \neq {\bf p_1}^{\prime}$, from the quantum superposition principle, contain information about the non-local or quantum behaviour of the system.

There is also an information theoretic two-electron uncertainty relationship 
\cite{guevarajcp1}
\begin{equation}
 0 \leq 6(1+ \ln \pi) \leq S_{\Gamma}^{u}+ S_{\Pi}^{u}=S_{T}^{u},  
\label{bial1001}
\end{equation}
where $S_{\Gamma}^{u}$ and $S_{\Pi}^{u}$ are unity-normalized two-electron Shannon entropies in the respective spaces, defined as
\begin{equation}
S_{\Gamma}^{u}= - \int \frac{\Gamma({\bf{r}}_1, {\bf{r}}_2)}{N(N-1)} \ln {\Bigg [}\frac{\Gamma({\bf{r}}_1, {\bf{r}}_2)}{N(N-1)}{\Bigg ]} d{\bf{r}}_1 d{\bf{r}}_2,
\label{shanr1}
\end{equation}
\begin{equation}
S_{\Pi}^{u}= - \int \frac{\Pi({\bf{p}}_1, {\bf{p}}_2)}{N(N-1)} \ln {\Bigg [}\frac{\Pi({\bf{p}}_1, {\bf{p}}_2)}{N(N-1)}{\Bigg ]} d{\bf{p}}_1 d{\bf{p}}_2.
\label{shanp1}
\end{equation}
$\Gamma({\bf{r}}_1, {\bf{r}}_2)$ and
$\Pi({\bf{p}}_1, {\bf{p}}_2)$
are the spinless two-electron densities, in position and
momentum space respectively, normalized to $N(N-1)$. These densities are the diagonal elements of the spin-free two-electron reduced density matrices, 
%
$\Gamma({\bf{r}}_1, {\bf{r}}_2)= \Gamma ({\bf r}_1,{\bf r}_2;{\bf r}_1,{\bf r}_2)$ and
$\Pi({\bf{p}}_1, {\bf{p}}_2)= \Pi({\bf p}_1,{\bf p}_2;{\bf p}_1,{\bf p}_2)$, where
\begin{eqnarray}
\Gamma ({\bf r}_1,{\bf r}_2;{\bf r}_1^{\prime},{\bf r}_2^{\prime}) = N(N-1) \int 
\Psi ({\bf x}_1, {\bf x}_2, \cdots,{\bf x}_N)
\Psi^* ({\bf x}_1^{\prime}, {\bf x}_2^{\prime}, \cdots,{\bf x}_N)
\nonumber
\\
\delta(\sigma _1 - \sigma^{\prime} _1) \delta(\sigma _2 - \sigma^{\prime} _2) 
d{\sigma}_1 d{\sigma}_2 d{\sigma^{\prime}}_1 d{\sigma^{\prime}}_2d{\bf x}_3 \cdots d{\bf x}_N
\label{ODM2r}
\end{eqnarray}
\begin{eqnarray}
\Pi({\bf p}_1,{\bf p}_2;{\bf p}_1^{\prime},{\bf p}_2^{\prime}) = N(N-1) \int 
\Phi ({\bf y}_1, {\bf y}_2, \cdots,{\bf y}_N)
\Phi^* ({\bf y}_1^{\prime}, {\bf y}_2^{\prime}, \cdots,{\bf y}_N)
\nonumber
\\
\delta(\sigma _1 - \sigma^{\prime} _1) \delta(\sigma _2 - \sigma^{\prime} _2) 
d{\sigma}_1 d{\sigma}_2 d{\sigma^{\prime}}_1 d{\sigma^{\prime}}_2d{\bf y}_3 \cdots d{\bf y}_N.
\label{ODM2p}
\end{eqnarray}

The off-diagonal elements of the two-electron reduced density matrix in each space (${\bf r}_1 \not= {\bf r}_1^{\prime}, {\bf r}_2 \not={\bf r}_2^{\prime}$ and ${\bf p}_1 \not= {\bf p}_1^{\prime}, {\bf p}_2 \not= {\bf p}_2^{\prime}$) supplies information about the non-local behavior at the two-electron level. It is natural to examine these regions of the two-electron reduced density matrices if one is interested in examining how the quantum nature of an electronic system is determined or governed by electron correlation. One can appreciate that although the information is encoded in the one- and two- electron density matrices, it is not an easy task to extract it since these quantities are six and twelve dimensional respectively, in nature.

The localization-delocalization phenomenon, and its consequences,  has been investigated at the one-electron level by studying $S_{\rho}$ and $S_{\pi}$ 
in atomic and molecular systems upon varying the molecular geometry \cite{hoijqc} or basis set in an isoelectronic series \cite{hojpb,hocpl}, and in the progress of a chemical reaction \cite{hosn2}. The entropy sum, $S_{\rho}+ S_{\pi}$,  has been proposed as a measure of electron correlation \cite{guevarapra}, and has been examined in confined systems \cite{senjcp}. This sum and its components have been considered with regard to the complexity in atomic systems \cite{chatz}. $S_{\rho}$ has been reported to be discontinuous in the helium isoelectronic series \cite{kais1}, and the finite size scaling of the Shannon entropy has been discussed \cite{kais2}. At the two-electron level, $S_{\Gamma}$ and $S_{\Pi}$ have been analyzed in atomic systems \cite{guevarajcp1,amoxvilli,sagarjcp1,sagarjcp2}. Mutual information in both spaces, defined as, 
$I_r=2S_{\rho}-S_{\Gamma}$ and $I_p=2S_{\pi}-S_{\Pi}$, have been used to study electron correlation and localization \cite{sagarjcp1,sagarjcp2}. 
Atomic shells and electron localization in momentum space have also been examined \cite{kohout}.

It was suggested \cite{sagarjcp2} that the entropic uncertainty relationships [Eqs.(\ref{guevara03}) and (\ref{bial1001})] can be interpreted as measures of local and non-local information and that delocalization of the density in a particular representation, due to electron correlation effects, should translate into significant contributions from the off-diagonal elements. 
If this concept could be made more precise, one would be able to examine the restrictions imposed by the uncertainty principle, and the effects of electron correlation, on the off-diagonal elements. This would provide us with conceptual information on how electron correlation influences the quantum properties of an electronic system. 

\subsection{Structure Factors}

Atomic structure factors are well known quantities and have been studied intensively over the years in position space \cite{schmiderjcp,schmiderjpb,galvezijqc,chen} and in momentum space \cite{schmiderjpb,thakkarcp,romerajcp}. In momentum space, they are also known as reciprocal forms factors or internally folded densities \cite{thakkarcp}. 

The spherically averaged quantities are defined in terms of the one-electron reduced density matrices as

\begin{equation}
F(k)= (4\pi)^{-1}\int \hat{\gamma}({\bf p};{\bf p + k})d{\bf p}d{\Omega} _k
\label{fk}
\end{equation}
\begin{equation}
B(s)= (4\pi)^{-1}\int \gamma({\bf r};{\bf r + s})d{\bf r}d{\Omega} _s,
\label{bs}
\end{equation}
or by use of the Fourier convolution theorem, they can be related to the Fourier components of the respective densities,
\begin{equation}
F(k)= (4\pi)^{-1} \int \rho({\bf r})e^{i{\bf k} \cdot {\bf r}} d{\bf r}d{\Omega}_k
= 4\pi \int \rho (r) j_0 (kr)r^2 dr
\label{fkft}
\end{equation}
\begin{equation}
B(s)= (4\pi)^{-1} \int \pi({\bf p})e^{-i{\bf s} \cdot {\bf p}} d{\bf p}d{\Omega}_s
= 4\pi \int \pi (p) j_0 (sp)p^2 dr.
\label{bsft}
\end{equation}
$j_0$ is the zero-order spherical Bessel function and the last equality in the previous two equations is due to the spherical symmetry of the system. 
F$(k)$ is obtainable from elastic scattering experiments while B$(s)$ can be determined from Compton profiles.

F$(k)$ and B$(s)$ are also known as autocorrelation functions from Fourier transform theory, and the normalization F$(0)$=B$(0)$=N is normally used. We emphasize that autocorrelation is distinct from electron correlation. One question, which we will address, is: How does electron correlation influence or impact upon the autocorrelation? These quantities provide information about the (integrated) off-diagonal elements as a function of the distance from the diagonal, and hence the quantum behavior in each representation. Another interpretation is that they are contracted forms of the one-electron reduced density matrix \cite{howard}. For example, one would expect that a more rapidly decaying B$(s)$ curve  would be associated with a more classical behavior, i.e. there are smaller contributions from the off-diagonal regions at larger $s$. 
The local behavior of the F$(k)$ curve in atomic systems has been studied along with its Laplacian, $\nabla^2$F$(k)$ \cite{schmiderjcp,galvezijqc}. The behaviors of B$(s)$ \cite{thakkarcp,romerajcp,schmiderthesis} and  $\nabla^2$B$(s)$ \cite{schmiderthesis}
have also been reported. 

The shape of the F$(k)$ and B$(s)$ curves may also be interpreted as the range of the electrons in each space. A sharply decaying curve corresponds to a smaller range while a flatter curve to a larger range, i.e. a larger range corresponds to a larger spatial extension of the curve.  Related to this is the concept of localization-delocalization since a more localized system would involve an autocorrelation function which decays more rapidly while a delocalized system would correspond to a slowly decaying autocorrelation function, i.e. significant contributions to the off-diagonal elements of the density matrices. Such a situation is identified with the concept of off-diagonal long range order which can be described as $\gamma({\bf r};{\bf r+s}) \not= 0$ for ${\bf |s|}$ large \cite{yang}. We note that off-diagonal long range order in one- and two-particle density matrices has been linked to phenomena such as Bose-Einstein condensation and superconductivity.

The question is now: Could one quantify the extent of this order, by a simple numerical measure, and hence make its meaning more precise?  The word $order$ leads us to consider the F$(k)$ and B$(s)$ distributions, from which one may use information theory and define entropic quantities to measure the global order or structure inherent in these distributions. That is, we study the structure or order in the distributions, whose independent variables, $k$ and $s$, contain the quantum properties of the system. We remark that information theory is not the only tool available to study the structure in a distribution. One can think of others, e.g. variance measures, which may serve the same purpose. 

The Shannon entropy in information theory \cite{shannon}
[defined for the electronic one-electron densities in Eqs. (\ref{shanr}) and (\ref{shanp})] is a measure of the structure or uncertainty in the underlying density and hence its use in the entropic uncertainty relationship in Eq. (\ref{guevara03}). Another interpretation is that it measures the spread, or in other words, the range of the distribution. More structured or ordered densities are associated with smaller values of the entropy while  larger values with flatter, less-ordered densities. In this context, we emphasize that long range order would be equated with disorder in the information theoretic framework and larger entropic values. 

We thus define the entropic quantities $S_F$ and $S_B$,
\begin{equation}
S_F= -4 \pi \int F^u(k) \ln F^u(k) k^2 dk
\label{sfk}
\end{equation}
\begin{equation}
S_B= -4 \pi \int B^u(s) \ln B^u(s) s^2 ds
\label{sbs}
\end{equation}
as measures of the order-disorder in the F$(k)$ and B$(s)$ curves, for F$(k)$ and B$(s)$$\geq 0$. We will remark on this restriction later on in the paper. Long range order (slowly decaying curves) should be associated with larger values of $S_F$ and $S_B$ while smaller values with the emergence of a more classical behavior. These quantities may also be interpreted as a measure of the width of the autocorrelation functions. 
We use unity-normalized distributions throughout,
\begin{equation}
4\pi \int F^u(k) k^2 dk = 4 \pi \int B^u(s) s^2 ds = 1.
\label{normfb}
\end{equation}
%

The purpose of this paper is to quantify and to examine the order present in the off-diagonal elements of density matrices, by definition and study of  the Shannon entropies of atomic structure factors, in position and in momentum space. We are interested in how the uncertainty principle governs the behavior of the off-diagonal elements and the effects of electron correlation on the off-diagonal order. We use as our test systems hydrogenic and helium ground state isoelectronic series. Atomic units are used throughout the paper. 

\section{Results and Discussion}

\subsection{Hydrogenic systems}
We begin the discussion by examining hydrogenic systems, in part to illustrate the concepts outlined in the introduction. For these systems, F$(k)$ and B$(s)$ \cite{thakkaracp} take the forms
\begin{equation}
F(k) = \frac{ 16Z^4}{(4Z^2+k^2)^2} \hspace{2cm} B(s)=e^{-Zs}(1+Zs+\frac{Z^2s^2}{3})
\label{fkbshy}
\end{equation}
where $Z$ is the nuclear charge, from which
\begin{equation}
S_F= 2(1+\ln \pi) + 7 \ln 2 + 3 \ln Z
\label{sfkhy}
\end{equation}
and
\begin{equation}
S_B= 2(1+\frac{\ln \pi}{2}) + 6 \ln 2 - 3 \ln Z + c.
\label{sbshy}
\end{equation}
The value of c is $\approx 0.0368$ and is due in part to the dependence of the expression on the exponential integral. Considering
Eqs. (\ref{sfkhy}) and (\ref{sbshy}), one can appreciate that the Z-dependence on summing them would cancel and would thus be constant throughout the hydrogenic series.

We present in Figs. 1 and 2, plots of F$(k)$ and B$(s)$ for some members of the hydrogenic series to illustrate some points. One notes that increasing the nuclear charge makes B$(s)$ more structured, or it decays more rapidly. Thus, the off-diagonal elements of the density matrix are being supressed with larger $Z$. On the other hand, F$(k)$ decays more slowly with increasing $Z$ and one may say that the off-diagonal elements of the density matrix are thus more significant in the momentum representation, with larger $Z$. In the literature, one encounters such results described by the terminology coherence and decoherence, especially when applied to optical systems. Decoherence, in this context, is characterized by a more sharply decaying distribution and accompanied by a suppression of the off-diagonal elements. In the limit, as $Z \rightarrow \infty$, this decay becomes extreme. One should also notice the complementary behavior, i.e. more structured in one space corresponds to less structured in the other. 

We now assimilate the structure, or decay characteristics of these curves, into a single global measure, by calculating their entropies  defined in Eqs. (\ref{sfk}) and (\ref{sbs}). We plot in Fig. 3, $S_F$ and $S_B$ for the members of the hydrogenic series. One observes that $S_B$ diminishes with $Z$ (more ordered or structured) while $S_F$ increases (less ordered). This behavior reflects the nature of the curves in Figs. 1 and 2. Note also the inverse behavior between $S_F$ and $S_B$, that is, between the momentum and position space representations. 

The result that the sum $S_F+S_B$ is a constant for the entire hydrogenic series reflects a certain symmetry between $S_F$ and $S_B$ values. With increasing $Z$, the B$(s)$ curves are localized to the same extent that the F$(k)$ curves are delocalized. A constant entropy sum for the densities has also been observed in the hydrogenic series \cite{guevarapra}, whose value is less than the entropy sum of the structure factors.

It is also instructive to compare these results with those for the ground state of the one-dimensional harmonic oscillator. In such one-dimensional systems, F$(k)$ and B$(s)$ are
\begin{equation}
F(k) =  e^{-\frac{k^2}{4\omega}} \hspace{2cm} B(s)= e^{-\frac{s^2\omega}{4}}
\label{fkbsosc}
\end{equation}
where $\omega$ is the frequency of oscillation. The corresponding  unity-normalized entropies are
\begin{equation}
S_F= \frac{1}{2}(1+\ln \pi) + \ln 2 + \frac{1}{2} \ln \omega, \hspace{2cm} S_B=\frac{1}{2}(1+\ln \pi) + \ln 2 - \frac{1}{2} \ln \omega.
\label{sfkosc}
\end{equation}
Thus the entropy sum of the structure factors in these systems is also independent of the potential. Its value, $(1+\ln \pi) + 2 \ln 2$, can be compared and contrasted to the entropy sum of the densities, $(1+\ln \pi)$ \cite{bbm,gadre}.

The additional term, $2 \ln 2$, in the entropy sum of the structure factors, is due to the differences in the normalization constants of the densities and the structure factors. The harmonic oscillator wave function corresponds to the minimum uncertainty one, that is, the lower bound of the uncertainty relationship for the Shannon entropies of the densities. Thus atomic systems with Coulomb-type potentials have entropy sums larger than the lower bound of the three-dimensional oscillator \cite{guevarajcp1,yanez}. On comparing the entropy sum of structure factors in the hydrogenic case to that of  the (three-dimensional) harmonic oscillator, one observes that the hydrogenic value is also larger than that of the oscillator. This suggests that the entropy sum of structure factors for the oscillator forms the lower bound for all systems.

\subsection{Interacting systems: The helium isoelectronic series}
We next consider the impact of electron correlation by examining members of the helium isoelectronic series. Simple model wave functions may be written for the ground states of these singlet systems as
\begin{equation}
\Psi_S(r_1,r_2) = C_N(e^{-Z_1r_1}e^{-Z_2r_2} + e^{-Z_2r_1}e^{-Z_1r_2}),
\label{hewfns}
\end{equation}
where $C_N$ is the normalization constant and $Z_1$ and $Z_2$ are variational parameters. F$(k)$ and B$(s)$ were calculated by spherical Bessel transforming the respective charge and momentum densities. $S_F$ and $S_B$ were calculated by numerical integration. In this series, the nuclear charge can be thought of as the parameter which controls the electron correlation. The model wave functions in an isoelectronic series provide a systematic manner of varying the effects of electron correlation. As $Z$ increases, $Z_1 \rightarrow Z_2 \rightarrow Z$ and the correlation decreases in a monotonic manner. Thus, the more highly correlated members of the series are those with smaller $Z$. Electron correlation in this model is purely Coulombic since the spatial part of the wave functions is symmetric. 

One can also envisage  non-interacting (NI) reference systems by setting $Z_1=Z_2=Z$ in the above. Such hydrogen-like systems would possess no electron correlation or interaction between electrons  and would be a convenient reference from which one could gauge the effects of correlation on the studied properties. 

We present in Figs. 4 and 5 plots of B$(s)$ and F$(k)$  for the helium series. One finds that the B$(s)$ curve corresponding to the helium member, the most correlated one, is the most diffuse one. With larger $Z$, the curve decays more sharply as the electron correlation is turned off. The  corresponding NI curve for helium decays more sharply than the interacting one. Thus, the effect of electron correlation is to make the B$(s)$ more diffuse. Correlation, in position space, is accompanied by more significant contributions to the off-diagonal elements, or delocalization of the Fourier components of the momentum density.

The case of F$(k)$ is opposite to that of B$(s)$. The highly correlated (Z=2) member is characterized by the most localized curve. As $Z$ increases, the F$(k)$ curve becomes more delocalized or flatter.  On comparison to the NI curve, the effect of correlation is to make the F$(k)$ curve more localized or abruptly decaying. That is, correlation in momentum space decreases the contributions from the off-diagonal elements, i.e localization of the Fourier components of the charge density. One may say that correlation in momentum space forces information closer to the diagonal part of the density matrix, i.e. the momentum density. The effects of electron correlation on both B$(s)$ and F$(k)$ \cite{schmiderjpb}, and on F$(k)$ \cite{chen}, have been studied in the lithium isoelectronic series.

Next, we show in Fig. 6 the behavior of  $S_F$ and $S_B$ in the He series. $S_F$ increases with $Z$ while $S_B$ decreases, showing a complementary-type relationship between the two. The most highly correlated system ($Z$=2) exhibits the most long-range order (largest entropy) in position space and the smallest in momentum space.

In Fig. 7 we plot $S_F$+$S_B$ and also include the sum for hydrogenic systems which corresponds to the non-interacting systems. The most striking observation is that the curve for the He series is not a constant, as in the case for non-interacting systems. The effect of correlation thus breaks the symmetry between the increase in the $S_F$ values and the decrease in the $S_B$ ones as witnessed in the non-interacting (hydrogenic) systems. 

One may also obtain a measure of the effect of electron correlation by defining
\begin{equation}
\Delta S_F= S_F - S_F^H  \hspace{2cm} \Delta S_B=S_B-S_B^H
\label{delta}
\end{equation}
where the superscript $H$ denotes the hydrogenlike value. Our results for all the studied members of the He series show that $\Delta S_B > 0$ while $\Delta S_F < 0$ and approach zero from above(below) as Z increases. Electron correlation induces a flattening  of the B$(s)$ curves  and a contraction of the F$(k)$ curves.
Thus, electron correlation can be thought of as a coherent phenomena (larger autocorrelation for larger values of $s$) in $r$-space, while it is a decoherent one in $p$-space, forcing the information into regions closer to the diagonal. 

\subsection{Two-electron atomic structure factors}
One can also consider two-electron structure factors \cite{thakkarcp} whose spherical averages are defined as
\begin{eqnarray}
F(k_1,k_2)= (4\pi)^{-2} \int \Pi ({\bf p}_1,{\bf p}_2;{\bf p}_1+{\bf k}_1, {\bf p}_2+{\bf k}_2) d{\bf p}_1 d{\bf p}_2 d\Omega_{k_1} d\Omega_{k_2}
\label{twofk}
\nonumber
\\
%
 = (4\pi)^{-2} \int \Gamma ({\bf r}_1,{\bf r}_2)e^{i [{\bf k}_1 \cdot {\bf r}_1+ {\bf k}_2 \cdot {\bf r}_2]} d{\bf r}_1 d{\bf r}_2 d\Omega_{k_1} d\Omega_{k_2}
\label{twofk1}
\end{eqnarray} 
and
\begin{eqnarray}
B(s_1,s_2)= (4\pi)^{-2} \int \Gamma ({\bf r}_1,{\bf r}_2;{\bf r}_1+{\bf s}_1, {\bf r}_2+{\bf s}_2) d{\bf r}_1 d{\bf r}_2 d\Omega_{s_1} d\Omega_{s_2}
\label{twobs}
\nonumber
\\
%
 = (4\pi)^{-2} \int \Pi ({\bf p}_1,{\bf p}_2)e^{-i [{\bf s}_1 \cdot {\bf p}_1+ {\bf s}_2 \cdot {\bf p}_2]} d{\bf p}_1 d{\bf p}_2 d\Omega_{s_1} d\Omega_{s_2}.
\label{twobs1}
\end{eqnarray} 
Equivalently,
\begin{equation}
F(k_1,k_2)= (4\pi)^{2} \int H(r_1,k_2)r_1^2 j_0(k_1r_1)dr_1,
\hspace{0.75 cm}
H(r_1,k_2)= \int \Gamma (r_1,r_2)r_2^2j_0(k_2 r_2)dr_2
\label{fktwosph}
\end{equation}
\begin{equation}
B(s_1,s_2)= (4\pi)^{2} \int K(p_1,s_2)p_1^2 j_0(s_1p_1)dp_1,
\hspace{0.75 cm}
K(p_1,s_2)= \int \Pi (p_1,p_2)p_2^2j_0(s_2 p_2)dp_2.
\label{bstwosph}
\end{equation}

These quantities are two-electron autocorrelation functions  and provide information about the range as a function of two electrons. The usual normalization is that $F(0,0)$=$B(0,0)$=${\genfrac{(}{)}{0pt}{}{N}{2}}$.
Electron correlation effects on the quantum properties, or range, would be contained in the correlation between the Fourier components of the respective densities, that is, between $k_1$ and $k_2$, and $s_1$ and $s_2$. To our knowledge, the behavior of atomic two-electron autocorrelation functions has not been studied in the literature.

We show their behavior in Figs. 8 and 9 for the helium atom. Note that the information in the two-electron density matrix of this system resides in regions relatively close to the diagonal, i.e. B($s_1$,$s_2$) and F($k_1$,$k_2$) are small for $s_1$,$s_2$ and $k_1$,$k_2$ not very large. The same general trends that were observed at the one-electron level are present.  That is, F($k_1$,$k_2$) decays more slowly than B($s_1$,$s_2$). With increasing $Z$, B($s_1$,$s_2$) decays more abruptly while F($k_1$,$k_2$) decays more slowly. This is illustrated with Be$^{+2}$ in Figs. 10 and 11.

Entropies for these distributions are defined as
\begin{equation}
S_F^2 = -(4\pi)^2 \int k_1^2 k_2^2 F^u(k_1,k_2) \ln F^u(k_1,k_2) dk_1 dk_2
\label{sfktwo}
\end{equation}
\begin{equation}
S_B^2 = -(4\pi)^2 \int s_1^2 s_2^2 B^u(s_1,s_2) \ln B^u(s_1,s_2) ds_1 ds_2
\label{sbktwo}
\end{equation}
where the two-electron structure factors are normalized to unity,
\begin{equation}
(4\pi)^2 \int F^u(k_1,k_2)k_1^2k_2^2dk_1dk_2=(4\pi)^2 \int B^u(s_1,s_2)s_1^2s_2^2ds_1ds_2=1.
\label{normtwo}
\end{equation}
These two-electron entropies provide a measure of the decay in the underlying distributions.
Two-electron structure factors were calculated from Eqns. (\ref{fktwosph}) and (\ref{bstwosph}) and their respective entropies from numerical integration of Eqns. (\ref{sfktwo}) and (\ref{sbktwo}).

One should also consider that in applying the definitions of the one- and two-electron entropies of structure factors, the presence of the logarithm demands that the underlying distributions be positive definite if one wishes the values of the entropies to be real numbers. We numerically verified this to be true for the studied systems. There are however, some atomic systems, which have negative regions in their B($s$) plots \cite{romerajcp} at the Hartree-Fock level of calculation. In such instances, one would have to employ other measures of the spread of a distribution, or re-define the distributions to be positive.

We plot in Fig. 12 $S_F^2$ and $S_B^2$ where the same general trends as those at the one-electron level are observed: $S_F^2$ increases (disordered) and $S_B^2$ decreases (ordered) with increasing Z. In Fig. 13, we plot the sum, $S_F^2+S_B^2$, along with the values for the corresponding non-interacting systems, $2(S_F^H + S_B^H)$. One can see that the difference between the two is largest for the most correlated (Z=2) member.  
The quantities $S_B^2-2S_B^H$ and $S_F^2-2S_F^H$ may be used as measures of how the range of the electrons is affected by the presence of Coulomb interactions between them. 
Our results yielded that $S_B^2-2S_B^H >0$ and $S_F^2-2S_F^H <0$ for all the studied members and approach zero as Z increases. Correlation is accompanied by delocalization in $r$-space and localization in $p$-space. 

\subsection{Information Distances}

One may also ask the question: How does electron correlation affect the correlation among the off-diagonal elements of the two-electron density matrix? Another way of phrasing the question is: 
How are the quantum properties of one electron correlated with those of the other electron? 
We examine this question by defining information distances
\begin{equation}
I_F= (4\pi)^2\int F^u(k_1,k_2) \ln {\Bigg [}\frac{F^u(k_1,k_2)}{F^u(k_1)F^u(k_2)}{\Bigg ]} k_1^2 k_2^2 dk_1dk_2 = 2S_F - S_F^2 \geq 0
\label{iff} 
\end{equation}
\begin{equation}
I_B= (4\pi)^2\int B^u(s_1,s_2) \ln {\Bigg [}\frac{B^u(s_1,s_2)}{B^u(s_1)B^u(s_2)}{\Bigg ]} s_1^2 s_2^2 ds_1ds_2 = 2S_B - S_B^2 \geq 0,
\label{ifb}
\end{equation}
which as off-diagonal correlation functions, provide one measure of the correlation between the variables $k_1$,$k_2$ and $s_1$,$s_2$.
Larger values imply more correlation between the variables.
These distances are measures of the interdependence or quantum interference effects that exist between two electrons in the system.

We show the behavior of $I_F$ and $I_B$ for the helium series in Fig. 14. One discovers that their behavior is similar, but that $I_B >I_F$, i.e. there is more interference between electrons in position space than in momentum space. Also, the Z=2 member possesses the largest correlation between variables which agrees with the argument that electron correlation is largest for this system. Another interpretation is that the largest correlation between variables in $r$-space, which may be taken as the extent of pairing between electrons, corresponds to the system with the largest entropy, or that with the most off-diagonal order. 



\begin{figure}[tb]
\begin{center}
   \includegraphics*[width=4.in,angle=0]{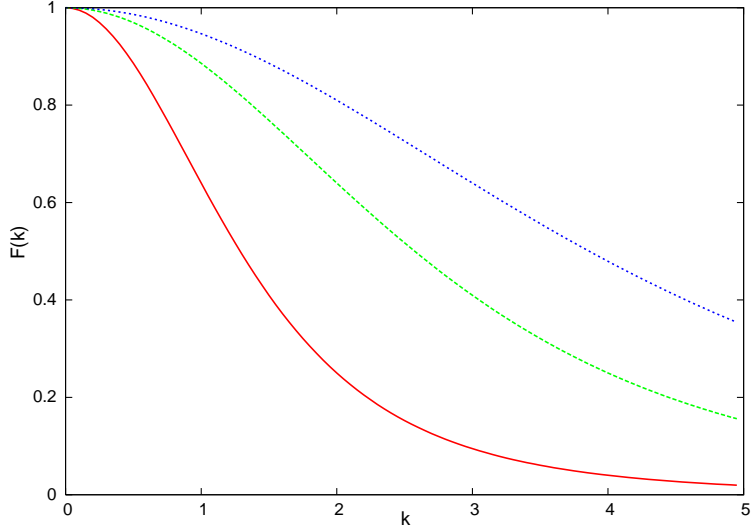}
    \caption{\label{fig1a} Plot of F$(k)$  for some members of the hydrogenic series: Z=2 (solid, red), Z=3 (dash, green), Z=4 (dot, blue).}
\end{center}
\end{figure}

\begin{figure}[tb]
\begin{center}
   \includegraphics*[width=4.in,angle=0]{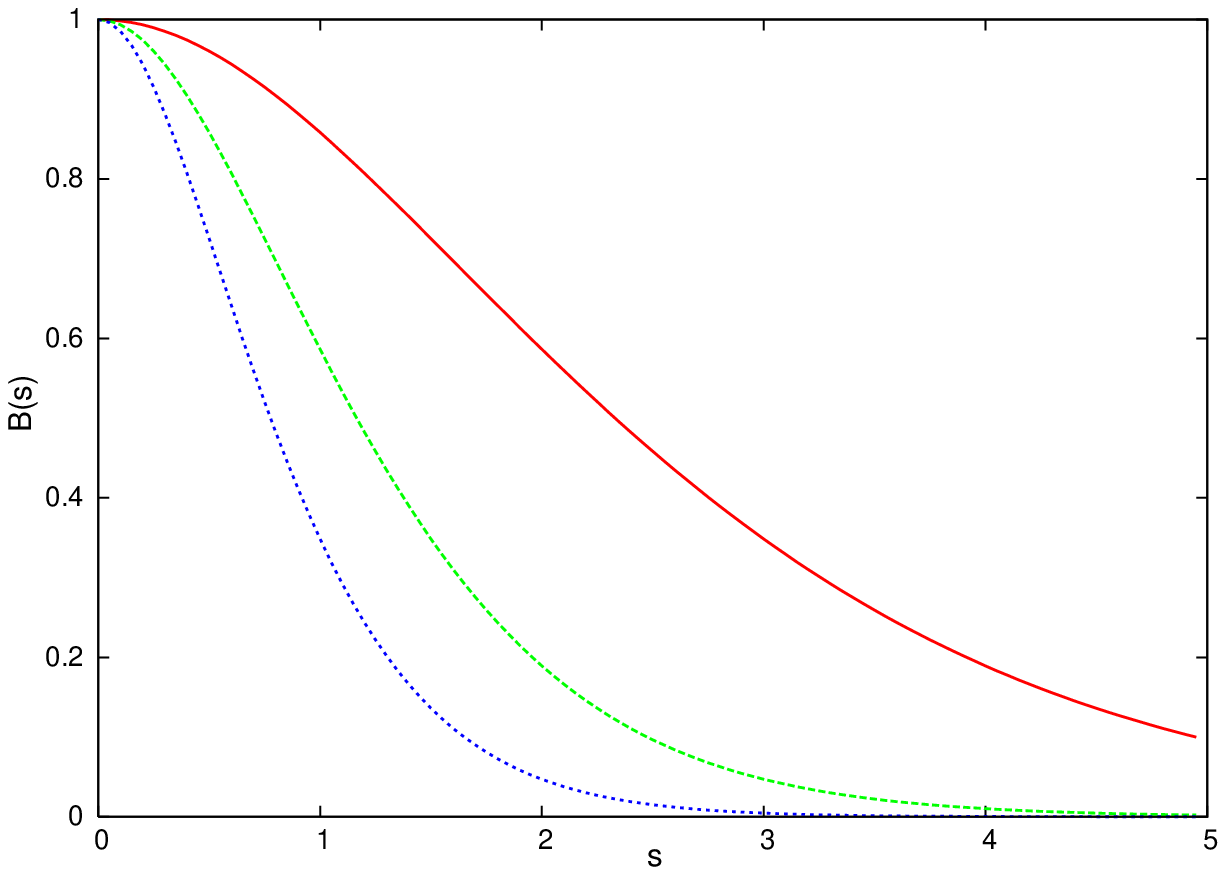}
    \caption{\label{fig1b}Plot of B$(s)$ for some members of the hydrogenic series: Z=2 (solid, red), Z=3 (dash, green), Z=4 (dot, blue).}
\end{center}
\end{figure}

\begin{figure}[tb]
\begin{center}
   \includegraphics*[width=4.in,angle=0]{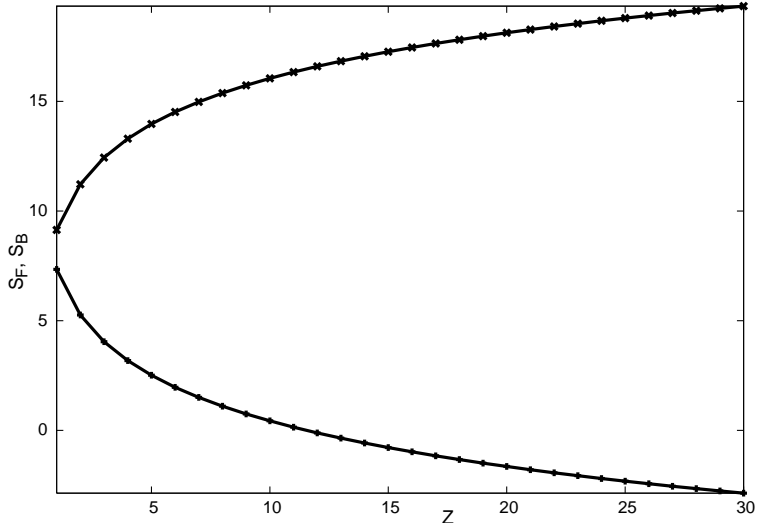}
    \caption{\label{fig2}$S_F$ (x's) and $S_B$ (pluses) for some members of the hydrogenic series $1 \leq Z \leq 30$.}
\end{center}
\end{figure}

\begin{figure}[tb]
\begin{center}
   \includegraphics*[width=4.in,angle=0]{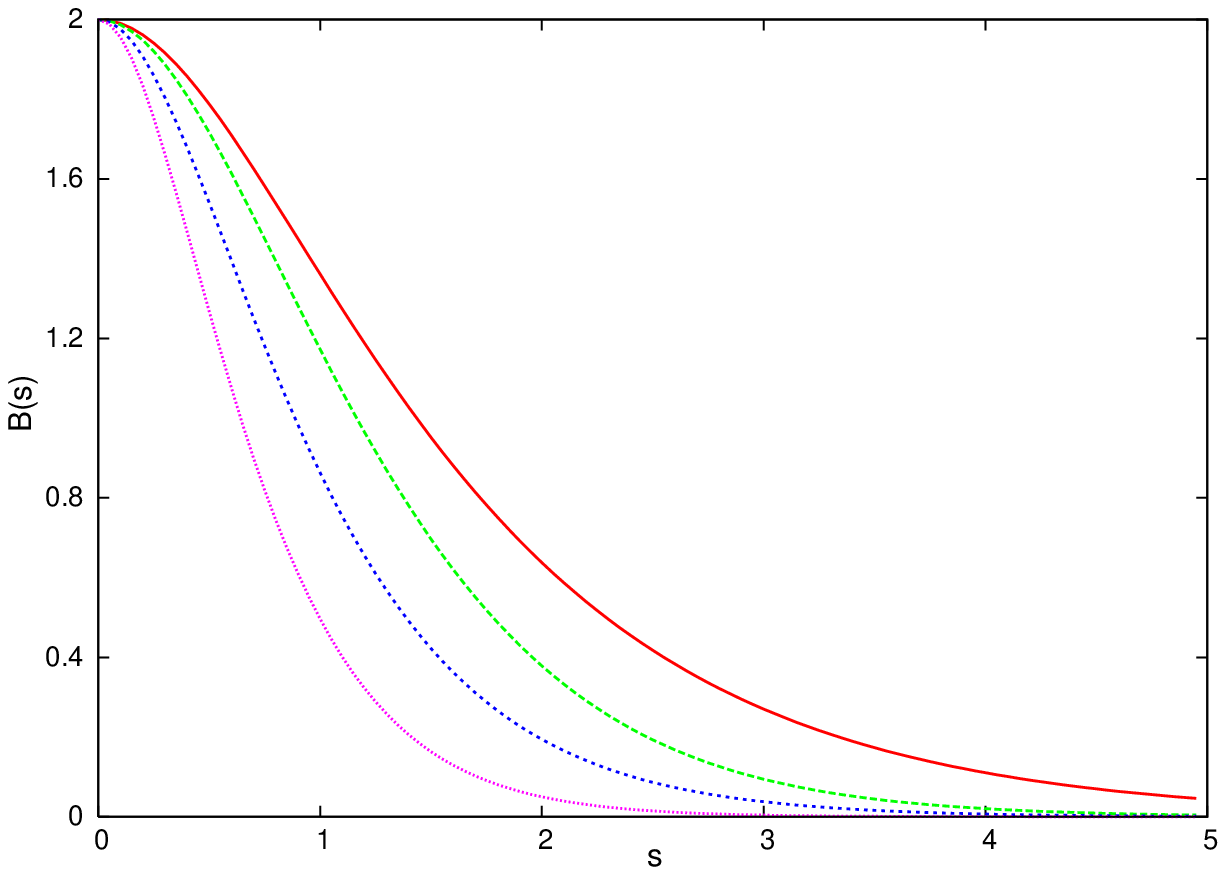}
    \caption{\label{fig3} Plot of B$(s)$ for some members of the helium series: Z=2 (solid, red), NI(Z=2) (dash,green), Z=3 (dot, blue), Z=4 (dash-dot, magenta).}
\end{center}
\end{figure}

\begin{figure}[tb]
\begin{center}
   \includegraphics*[width=4.in,angle=0]{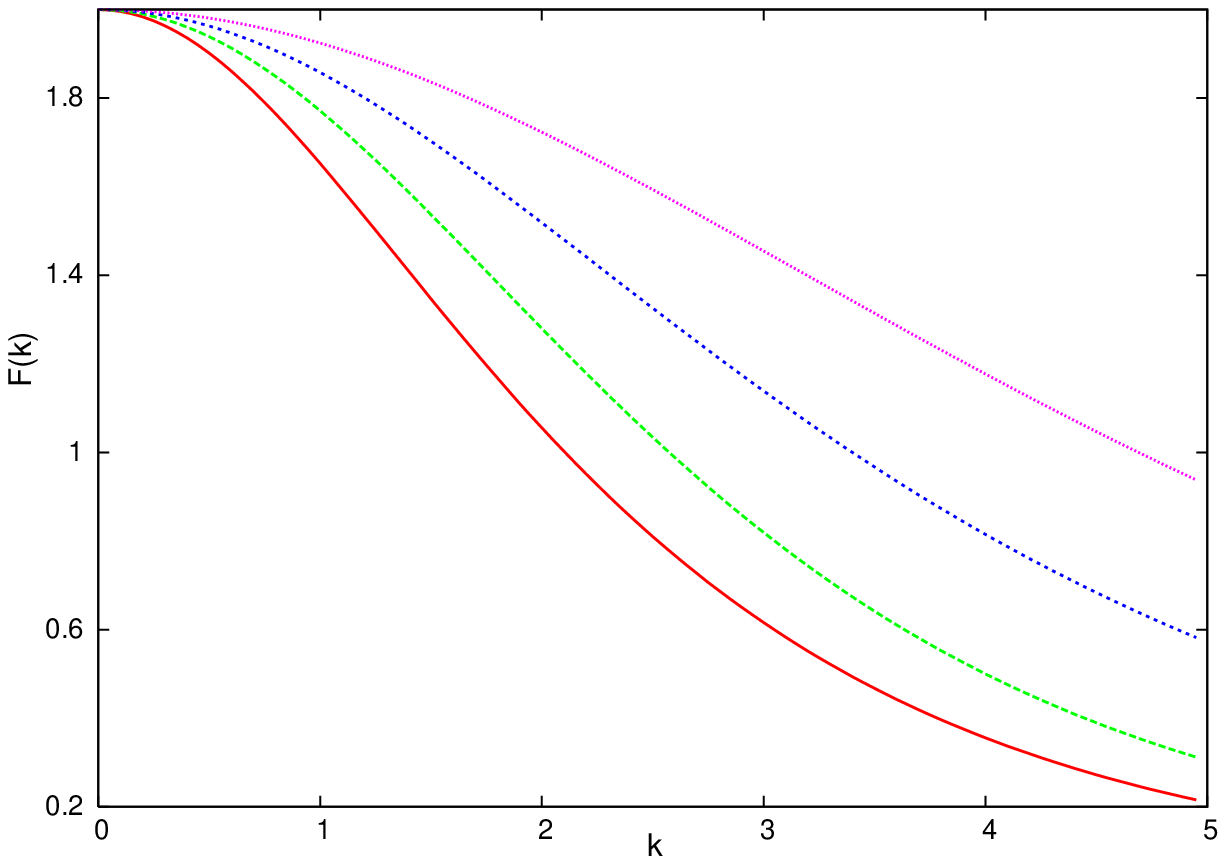}
    \caption{\label{fig4} Plot of F$(k)$ for some members of the helium series: Z=2 (solid, red), NI(Z=2) (dash, green), Z=3 (dot, blue), Z=4 (dash-dot, magenta).}
\end{center}
\end{figure}

\begin{figure}[tb]
\begin{center}
   \includegraphics*[width=4.in,angle=0]{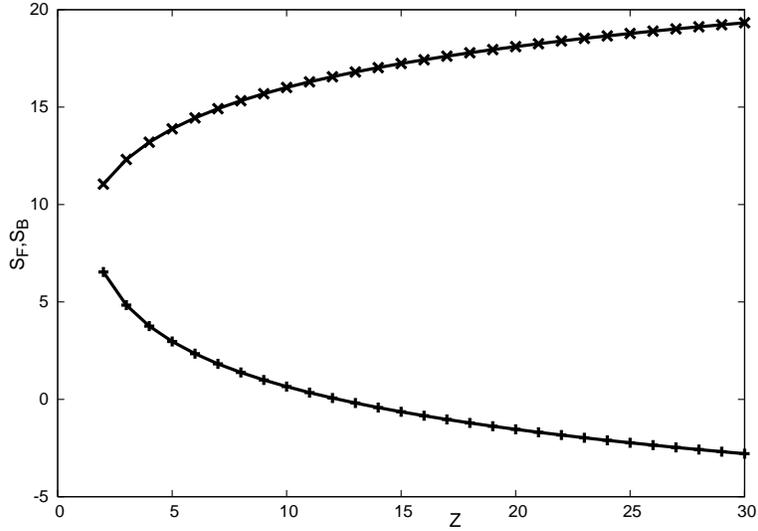}
    \caption{\label{fig5} $S_F$ (x's) and $S_B$ (pluses) for some members of the helium series.}
\end{center}
\end{figure}

\begin{figure}[tb]
\begin{center}
   \includegraphics*[width=4.in,angle=0]{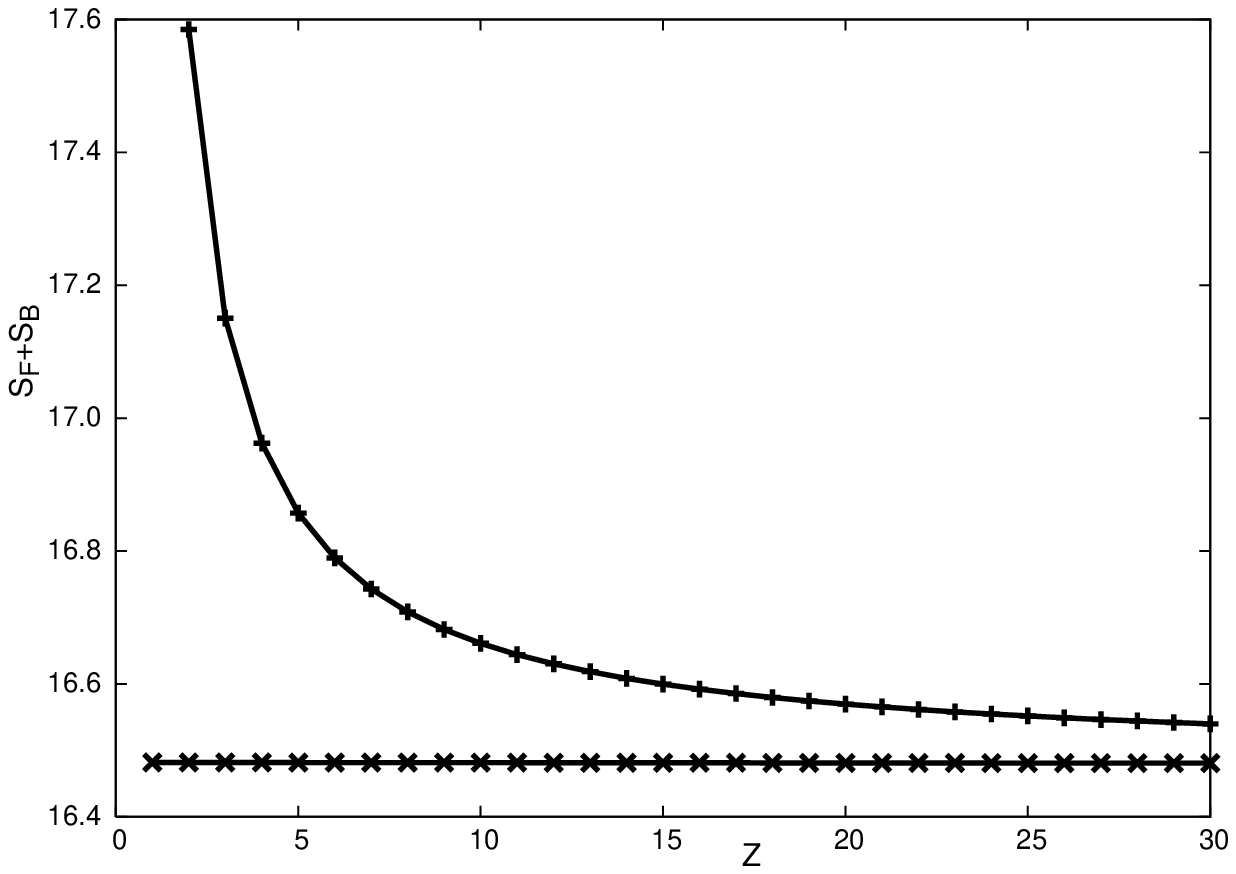}
    \caption{\label{fig6} $S_F+S_B$ for interacting (pluses) and non-interacting (x's) members of the helium series.}
\end{center}
\end{figure}

\begin{figure}[tb]
\begin{center}
   \includegraphics*[width=4.in,angle=0]{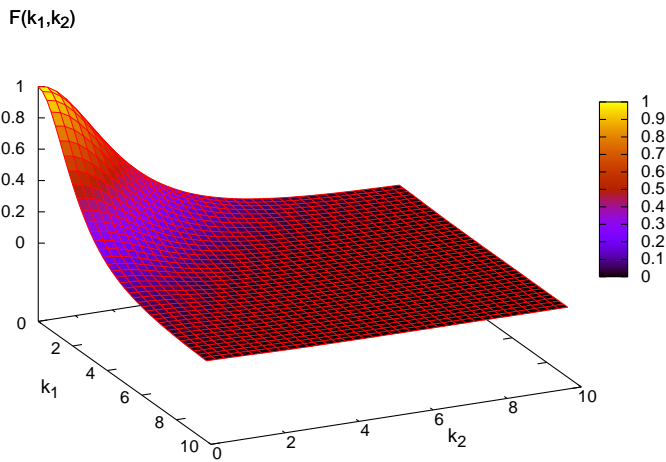}
    \caption{\label{fig7} Plot of F($k_1$,$k_2$) for the helium  atom.}
\end{center}
\end{figure}

\begin{figure}[tb]
\begin{center}
   \includegraphics*[width=4.in,angle=0]{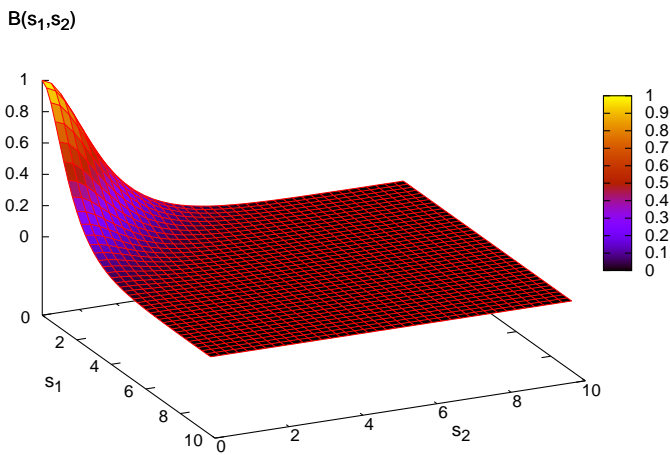}
    \caption{\label{fig8} Plot of B($s_1$,$s_2$) for the helium atom.}
\end{center}
\end{figure}

\begin{figure}[tb]
\begin{center}
   \includegraphics*[width=4.in,angle=0]{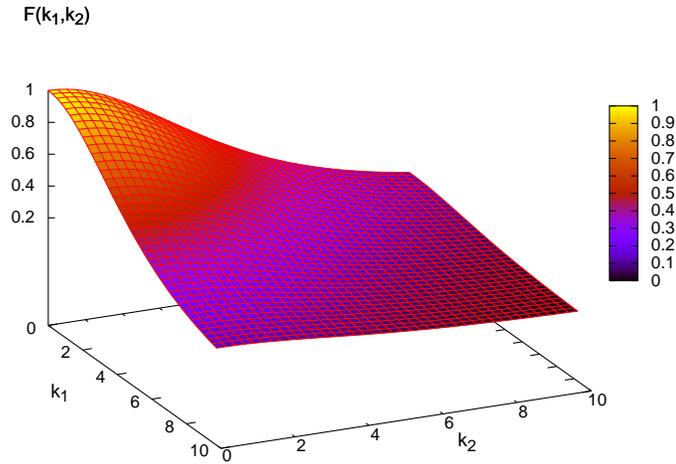}
    \caption{\label{fig9} Plot of F($k_1$,$k_2$) for the $Z=4$ member of the helium series.}
\end{center}
\end{figure}

\begin{figure}[tb]
\begin{center}
   \includegraphics*[width=4.in,angle=0]{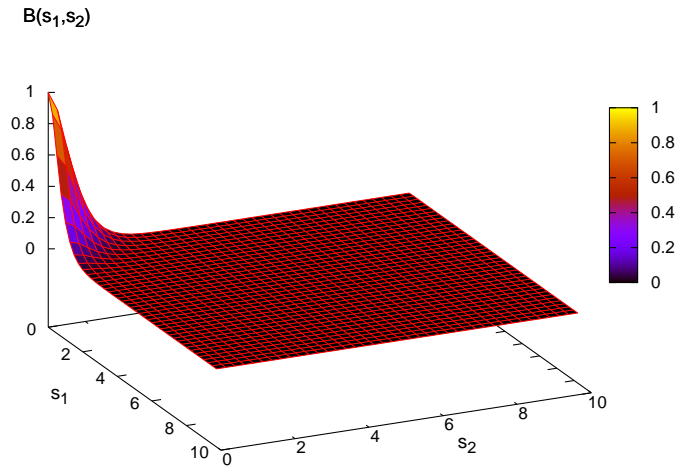}
    \caption{\label{fig10} Plot of B($s_1$,$s_2$) for the $Z=4$ member of the helium series.}
\end{center}
\end{figure}
\begin{figure}[tb]
\begin{center}
   \includegraphics*[width=4.in,angle=0]{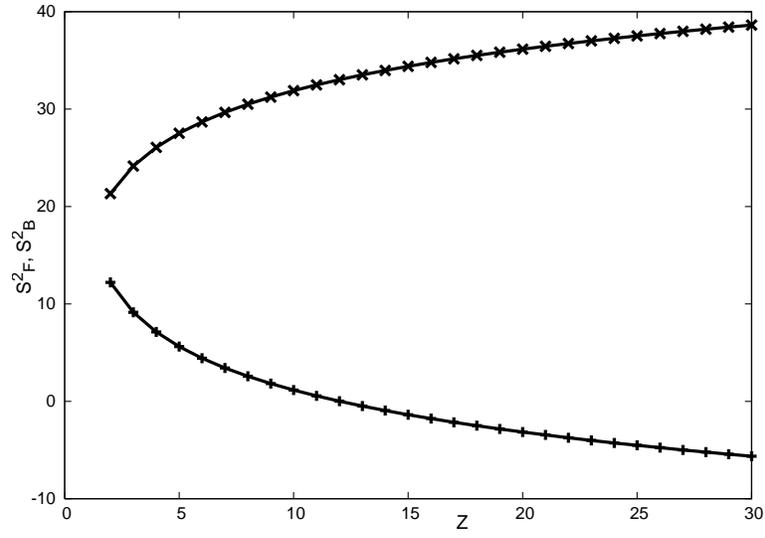}
    \caption{\label{fig11}$S_F^2$ (x's) and $S_B^2$ (pluses) for the helium series.}
\end{center}
\end{figure}

\begin{figure}[tb]
\begin{center}
   \includegraphics*[width=4.in,angle=0]{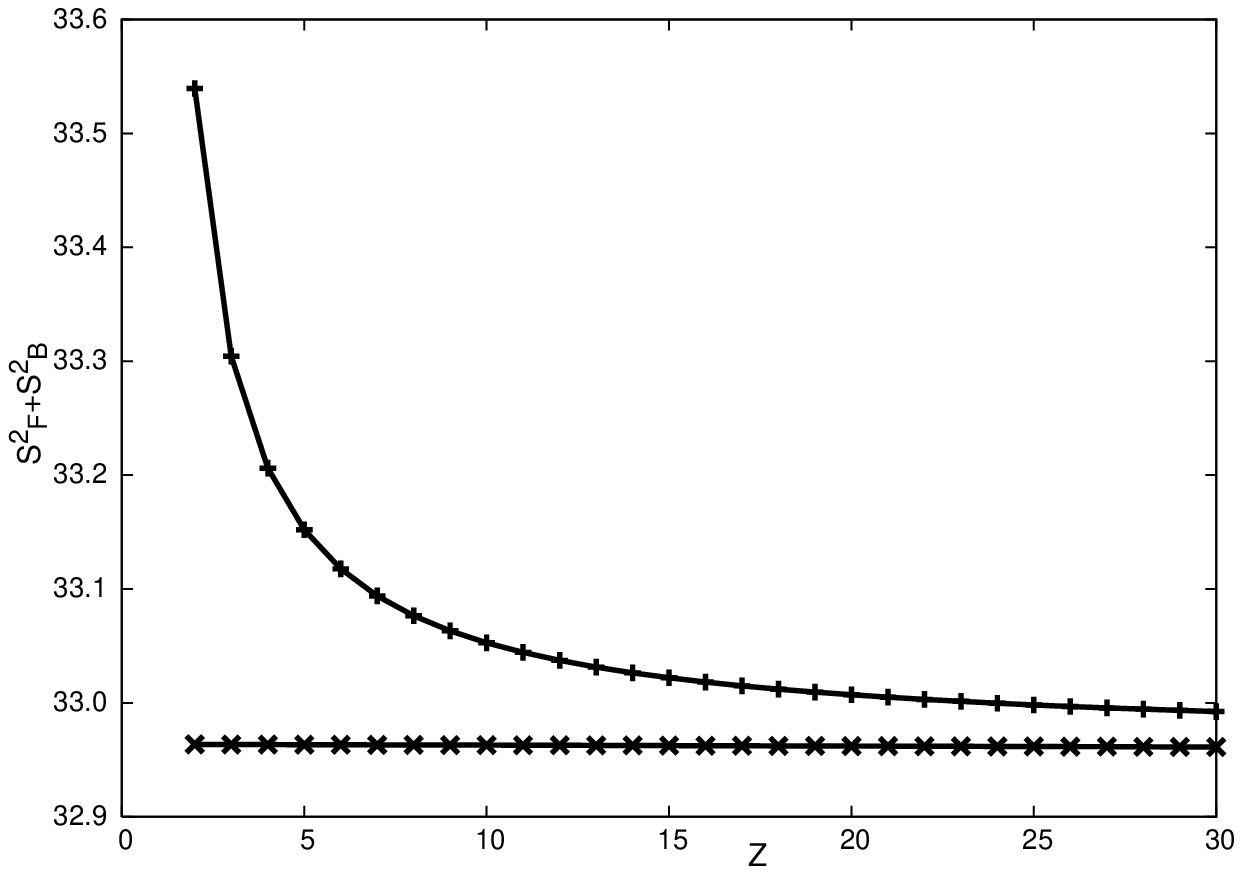}
    \caption{\label{fig12}$S_F^2+S_B^2$ for interacting (pluses) and non-interacting (x's) members of the helium series.}
\end{center}
\end{figure}

\begin{figure}[tb]
\begin{center}
   \includegraphics*[width=4.in,angle=0]{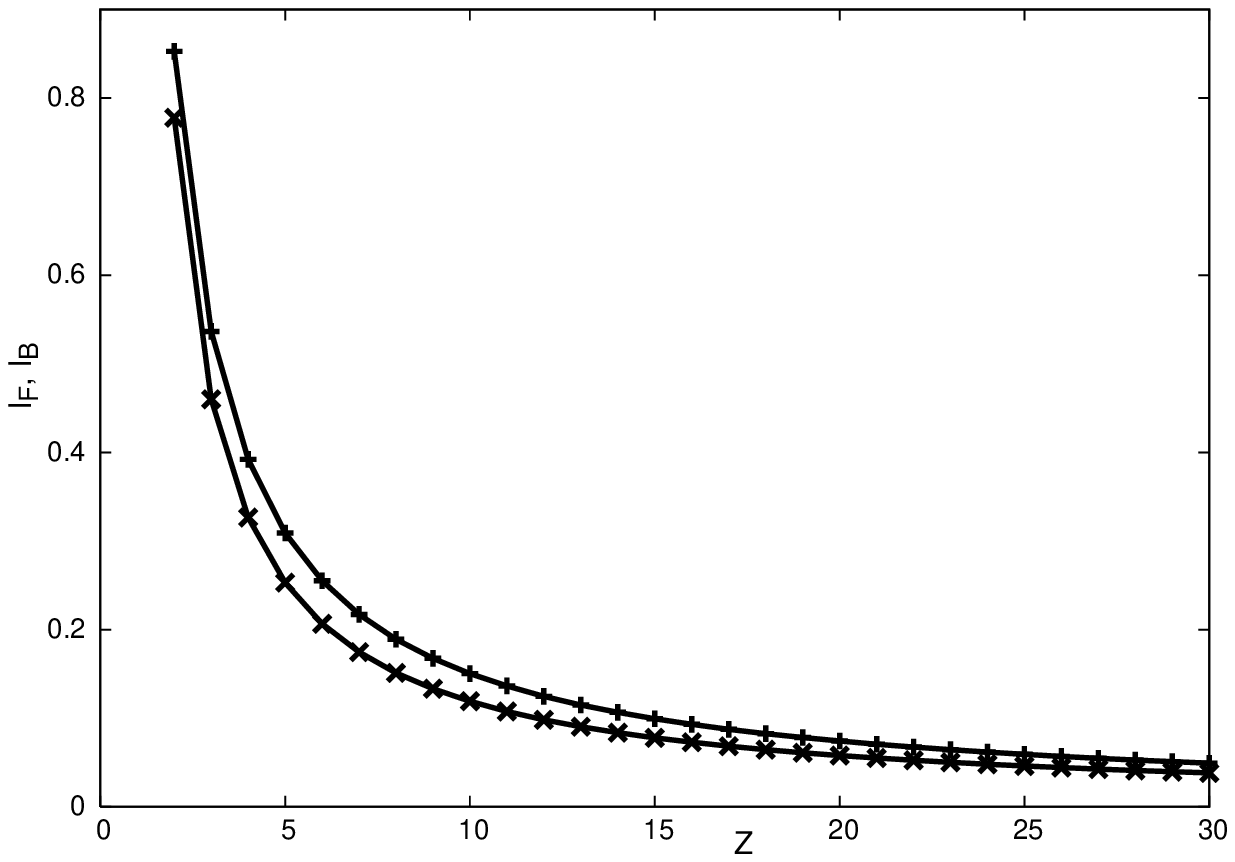}
    \caption{\label{fig13}$I_F$(x's) and $I_B$(pluses) for the helium series.}
\end{center}
\end{figure}

%


\section{Conclusions}
Shannon entropies of one- and two-electron atomic structure factors are used to examine off-diagonal order in one- and two-electron density matrices, in position and momentum space. The relationship between this order, the behavior of the structure factors and their entropies, and the uncertainty principle, is scrutinized in hydrogenic and helium-like systems. We show that the Shannon entropies of the structure factors obey a complementary-type relationship. We find that their sum is a constant in the ground state hydrogenic series and harmonic oscillator. That is, increasing order in the $r$-space structure factor is matched by an equivalent disorder in the $p$-space one. These quantities are then studied in the helium isoelectronic series to examine the effects of electron correlation. We observe that electron correlation induces a broadening in the $r$-space structure factor and a more sharply decaying $p$-space structure factor. This translates into the result that electron correlation causes a larger $r$-space entropy and a smaller $p$-space one. The interpretation of this is that electron correlation forces information into the off-diagonal elements of the one-particle density matrix in $r$-space, i.e. more off-diagonal order, while it forces information into regions closer to the diagonal in $p$-space. However, distinct from the hydrogenic case, the sum of the two entropies is not a constant as the symmetry between the increase and decrease is broken. This same type of behavior is also present for the two-electron structure factor entropies. The correlations between the variables, $k_1$ and $k_2$, $s_1$ and $s_2$, which carry the information about the quantum interference, is studied from information distances.  We notice that this correlation between variables is largest for the helium member. It is also the largest between $s_1$ and $s_2$ which corresponds to the off-diagonal behavior in the position space representation.


\section{ACKNOWLEDGMENTS}
The authors thank  the Consejo
Nacional de Ciencias y Tecnologia (CONACyt) and the PROMEP program of the Secretario de Educaci\'{o}n P\'{u}blica in M\'{e}xico for support. 

%
%



\begin{thebibliography}{30}

\bibitem{bbm} I. Bialynicki-Birula and J. Mycielski 1975 {\it Commun. Math. Phys.} {\bf 44} 129.

\bibitem{gadre} S.~R. Gadre, S.~B. Sears, S.~J. Chakravorty and R.~D. Bendale 1985 {\it Phys. Rev. A}
 {\bf 32} 2602.

\bibitem{guevarajcp1} N. L. Guevara, R. P. Sagar and R. O. Esquivel 2003 {\it J. Chem. Phys.}  {\bf 119} 7030.

\bibitem{davidson} E.R. Davidson 1976 {\it Reduced Density Matrices in Quantum Chemistry} (Academic: New York).


\bibitem{hoijqc} M. H{\^o}, R.P. Sagar, D.F. Weaver and V.H. {Smith, Jr.} 1995 {\it Int. J. Quantum Chem.} S{\bf 29} 109.

\bibitem{hojpb}
M. H{\^o}, R.~P. Sagar, V.~H. {Smith, Jr.} and R.~O. Esquivel 1994 {\it J. Phys. B.}
{\bf 27} 5149.

\bibitem{hocpl}
M. H{\^o}, R.P. Sagar, J.M. Perez-Jorda, V.~H. {Smith, Jr.} and R.~O. Esquivel 1994
{\it Chem. Phys. Lett.} {\bf 219}  15.

\bibitem{hosn2} M. H{\^o}, H. Schmider, D.F. Weaver, V.H. {Smith, Jr.}, R.P. Sagar and R.O. Esquivel 2000
{\it Int. J. Quantum Chem.} {\bf 77} 376.

\bibitem{guevarapra}  N. L. Guevara, R. P. Sagar and R. O. Esquivel 2003 {\it Phys. Rev. A} {\bf 67} 012507.

\bibitem{senjcp} K. D. Sen 2005 {\it J. Chem. Phys.} {\bf 123} 074110.

\bibitem{chatz} K. Ch. Chatzisavvas, Ch. C. Moustakidis and C. P. Panos 2005 {\it J. Chem. Phys.} {\bf 123} 174111.

\bibitem{kais1} Q. Shi and S. Kais 2005 {\it Chem. Phys.} {\bf 309} 127.

\bibitem{kais2} Q. Shi and S. Kais 2004 {\it J. Chem. Phys.} {\bf 121} 5611.

\bibitem{amoxvilli} C. Amovilli and N. H. March 2004 {\it Phys. Rev. A} {\bf 69} 054302.

\bibitem{sagarjcp1} R.P. Sagar and N.L. Guevara 2005 {\it J. Chem. Phys.} {\bf 123} 044108.

\bibitem{sagarjcp2} R.P. Sagar and N.L. Guevara 2006 {\it J. Chem. Phys.} {\bf 124} 134101.

\bibitem{kohout} M. Kohout, F.R. Wagner and Y. Grin 2006 {\it Int. J. Quantum Chem.} {\bf 106} 1499.
 
\bibitem{howard} I.A. Howard, N.H. March and V.E. Van Doren 2001 {\it Phys. Rev. A} {\bf 64} 042509.

\bibitem{schmiderjcp} H. Schmider, R.P. Sagar and V.H. Smith, Jr. 1991 {\it J. Chem. Phys.} {\bf 94} 4346.

\bibitem{schmiderjpb} H. Schmider, R.O. Esquivel, R.P. Sagar and V.H. Smith, Jr. 1993 {\it J. Phys. B.} {\bf 26} 2943.

\bibitem{galvezijqc} F.J. G{\'a}lvez and I. Porras 1995 {\it Int. J. Quantum Chem.} {\bf 56} 157.

\bibitem{chen} C. Chen and Z.-W. Wang 2005 {\it J. Chem. Phys.} {\bf 122} 024305.

\bibitem{thakkarcp} A.J. Thakkar, A.M. Simas and V.H. Smith, Jr. 1981 {\it Chem. Phys.} {\bf 63} 175.

\bibitem{romerajcp} E. Romera and J.C. Angulo 2004 {\it J. Chem. Phys.} {\bf 120} 7369.

\bibitem{schmiderthesis} H. Schmider 1994 Ph.D. thesis, Queen's University at Kingston.

\bibitem{yang} C.N. Yang 1962 {\it Rev. Mod. Phys.} {\bf 34} 694.

\bibitem{shannon}
C.~E. Shannon 1948 {\it Bell Syst. Tech. J.} {\bf 27} 379, reprinted in
1993  \textit{Claude Elwood Shannon: collected papers}(IEEE Press: New York).


\bibitem{thakkaracp} A.J. Thakkar 2003 {\it Adv. Chem. Phys.} {\bf 128} 303.

\bibitem{yanez} R.J. Y\'{a}\~{n}ez, W. Van Assche and J. S. Dehesa 1994 {\it Phys. Rev. A} {\bf 50} 3065.


\end{thebibliography}
\end{document}